\documentclass[sigconf]{acmart}

\setcopyright{none}
\renewcommand\footnotetextcopyrightpermission[1]{}
\acmConference{}{}{}
\acmDOI{}
\acmISBN{}

\usepackage{booktabs}       
\usepackage{tabularx}       
\usepackage{algorithm}      
\usepackage{algpseudocode}  
\usepackage{multirow}       
\usepackage{xspace}         
\usepackage{balance}        
\usepackage{orcidlink}      

\AtBeginEnvironment{tabular}{\small}
\AtBeginEnvironment{tabular*}{\small}

\newcommand{\civic}{\texttt{CIVIC\_ADDRESS}\xspace}
\newcommand{\noresult}{\texttt{NO\_RESULT}\xspace}
\newcommand{\loctype}{\texttt{location\_type}\xspace}
\newcommand{\rooftop}{\texttt{ROOFTOP}\xspace}
\newcommand{\resulttype}{\texttt{resultType}\xspace}
\newcommand{\housenum}{\texttt{houseNumber}\xspace}

\begin{document}

\title{Wrong yet Confident: Automated Detection of Silent Failures\\in Location-Dependent Software}

\author{Juan A. Padilla \orcidlink{0000-0002-2586-7802}}
\orcid{0000-0002-2586-7802}
\affiliation{%
  \department{Computer Science and Engineering Department}
  \institution{University of Puerto Rico, Mayag\"{u}ez}
  \city{Mayag\"{u}ez}
  \state{PR}
  \country{USA}
}
\email{juan.padilla11@upr.edu}

\begin{abstract}
We report an experience paper applying round-trip offset---an automated
testing oracle for location-dependent software---at deployment scale
across 32,019 GPS delivery transactions, corresponding to 24,501 unique
coordinates, from a single USDA COVID-19 food assistance deployment in
Bayam\'on, Puerto Rico. The oracle tests each geocoder's implicit
spatial accuracy claim by measuring the deviation between an input
coordinate and the coordinate recovered after reverse-then-forward
geocoding, without requiring manual labeling of correctness. We
evaluated one community-contributed and two commercial geocoders
against this operationally validated corpus. The experience surfaces
two distinct failure classes, which we report with their respective
denominators because they are not comparable quantities. First, 2,441
of the 24,501 unique coordinates (10.0\%) are invisible to all three
platforms: an acknowledged failure with no resolution, and a property
of the corpus rather than of any one geocoder. Second, and
platform-specific: 10.0\% of the 13,221 coordinates Google resolves to
a civic address exhibit round-trip offsets exceeding 2~km---5.4\% of
the full corpus---returned with no error signal, against 0.5\% for the
second commercial platform and 0.04\% for the community geocoder.
Round-trip offset detects all large-offset failures observed in this
dataset while standard resolution-rate evaluation detects none. We
document an eight-mode fault taxonomy, a structured comparison of
detection methods, and concrete lessons for integrating automated
geocoder verification into CI/CD pipelines. The key lesson: resolution
rate measures availability; round-trip offset measures correctness. In
non-standard addressing environments these two metrics select for
opposite behaviors, and only the latter catches the failures that matter
for autonomous action.
\end{abstract}



\maketitle

\pagestyle{fancy}
\fancyhf{}
\fancyhead[LO]{Wrong yet Confident: Automated Detection of Silent Failures}
\fancyhead[RE]{J. A. Padilla}
\fancyfoot[C]{\thepage}
\renewcommand{\headrulewidth}{0.4pt}
\renewcommand{\footrulewidth}{0pt}

\fancypagestyle{firstpage}{%
  \fancyhf{}%
  \fancyhead[C]{\small Preprint, \today. Previously submitted to ASE 2026.}%
  \fancyfoot[C]{\thepage}%
  \renewcommand{\headrulewidth}{0pt}%
  \renewcommand{\footrulewidth}{0pt}%
}
\thispagestyle{firstpage}

\section{Introduction}

Software systems routinely delegate correctness decisions to external
infrastructure. A routing system trusts its geocoding API. An emergency
dispatch platform trusts that the address it resolved is the address the
caller is at. An autonomous delivery vehicle trusts that the coordinate
it was given corresponds to the household on the manifest. This trust is
structurally invisible: there is no interface contract, no error
channel, and no runtime verification. The geocoding API returns a result
type, a match-precision label, and an address string. The calling system
treats all three as verified.

A system with direct access to operationally validated GPS coordinates
does not need geocoding at all---the coordinate is the ground truth. The
geocoding problem arises when that coordinate is unavailable: when a
follow-up program receives only an address from a referral or registry,
when a legacy Computer-Aided Dispatch (CAD) system takes address input from an emergency caller,
when a logistics platform works from a delivery manifest rather than a
GPS pin, or when operational GPS data was collected by one program and
not preserved for the next. Round-trip offset is the verification layer
for systems that cannot avoid the address-to-coordinate translation. It
answers the question those systems cannot answer for themselves: is the
address I was given one that geocoding can resolve correctly?

This paper reports an experience applying round-trip offset as an
automated testing oracle for verifying external location infrastructure
in software systems. The oracle exploits a metamorphic relation inherent
in geocoding: reverse-then-forward geocoding of a valid coordinate
should recover that coordinate within an acceptable spatial tolerance.
Any violation is an automatically detected correctness failure without
requiring manual labeling of correctness. The technique requires only
API access and a GPS coordinate corpus---both available as byproducts of
normal system operation---and scales to tens of thousands of locations
without manual review.

We evaluate the oracle against 32,019 GPS delivery transactions,
corresponding to 24,501 unique coordinates, from a single USDA COVID-19
food assistance deployment in Bayam\'on, Puerto Rico---a
stress-test context where formal urbanization addressing coexists with
kilometer-marker rural addressing~\cite{census14}, and where mainstream geocoding
platforms were built on continental US conventions. This context is not
the subject of the paper; it is the evaluation instrument.

The central finding is that standard evaluation---resolution rate,
match-precision label, result type---is not only insufficient but
actively misleading. Every one of the 30 cases in this dataset with
offset $>10$~km was returned as a \civic result carrying its platform's
strongest match-precision label and no error field.
Resolution-rate evaluation detected zero of them. Round-trip offset
detected all 30 without requiring any external ground-truth label: the
offset magnitude itself provides sufficient evidence of correctness
violation.

The oracle is not specific to geocoding. It applies to any system that
satisfies two conditions: (1)~the system relies on an external API that
makes implicit correctness claims through a match-precision or
result-type signal, and (2)~there exists an input-output transformation
that enables metamorphic validation. Other candidate domains include
address normalization APIs, routing services, and map-matching
pipelines.

\noindent This experience yields the following findings and lessons:
\begin{itemize}
\item \textbf{A reusable automated oracle for location-dependent
  software:} round-trip offset, a property-based metamorphic technique
  that verifies geocoder correctness claims without requiring manual
  labeling of correctness, deployable in CI/CD pipelines against any GPS
  coordinate corpus.

\item \textbf{Deployment-scale evidence:} 24,501 unique validated
  delivery coordinates drawn from 32,019 delivery transactions, three
  geocoders, showing that round-trip offset detects
  all 30 large-offset cases ($>10$~km, operationally implausible as
  correct resolutions) that resolution-rate evaluation misses, and that
  the oracle's detection boundary is structurally robust---failures
  declare themselves by magnitude, not by marginal threshold proximity.

\item \textbf{A structured comparison of detection methods (lesson for
  practitioners):} resolution rate, match-precision label,
  multi-geocoder agreement, and round-trip offset evaluated against the
  same failure corpus, showing which classes each method detects and
  misses.

\item \textbf{A fault taxonomy grounded in operational experience:}
  an eight-mode fault taxonomy partitioning evenly into four
  software-solvable modes and four that require civic addressing
  infrastructure rather than better software, grounded in
  deployment-scale evidence.
\end{itemize}

The paper is organized as follows. Section~\ref{sec:oracle} describes
the round-trip offset oracle. Section~\ref{sec:related} situates the
work in related literature. Section~\ref{sec:setup} describes the
evaluation setup. Section~\ref{sec:results} presents detection results.
Section~\ref{sec:taxonomy} presents the fault taxonomy.
Section~\ref{sec:discussion} discusses generalizability, limitations,
and lessons. Section~\ref{sec:conclusion} concludes.

\section{The Round-Trip Offset Oracle}
\label{sec:oracle}

This section presents the round-trip offset oracle: its theoretical
grounding in the oracle problem and metamorphic testing, its formal
specification, a structured comparison against alternative detection
methods, the conditions under which it applies, and the sense in which
it constitutes a new oracle rather than an application of an existing
one.

\subsection{Oracle Design}

Location-dependent software systems consume geocoding API outputs as
authoritative assertions: a \civic result carrying a platform's
strongest match-precision label---Google's \loctype set to \rooftop,
HERE's \resulttype set to \housenum---implicitly claims that the
returned address corresponds to the input coordinate. This claim is
never verified by the calling system---there is no return channel for
correctness, only a return channel for availability. The oracle
problem~\cite{barr2015oracle} is
precisely this: how do we determine whether a program's output is
correct when no ground-truth specification is available at runtime?

Round-trip offset resolves this by exploiting a metamorphic
relation~\cite{chen1998metamorphic} inherent in geocoding. A
metamorphic relation is a necessary property of correct outputs: if the
system is working correctly, certain transformations of the input should
produce predictably related outputs. For geocoding, the relation is:
reverse-then-forward geocoding of a valid coordinate $c$ should recover
a coordinate $c'$ close to $c$. Formally, given a ground-truth
coordinate $c$, let $a = \mathit{reverse}(c)$ be the civic address
returned by geocoder $G$, and let $c' = \mathit{forward}(a)$ be the
coordinate obtained by re-geocoding that address. The round-trip offset
is $d(c,c')$ in meters, computed as the Haversine great-circle
distance. A small offset confirms the implicit claim; a large offset is
an automatically detected correctness violation. No human oracle is
required. The GPS coordinate itself is the specification.

This is not circular. The relation is non-trivial because it tests the
\emph{content} of the geocoder's output, not just its presence.
Resolution rate tests whether a result was returned; round-trip offset
tests whether the returned result is spatially consistent with the
input. A geocoder can achieve 100\% resolution rate while every result
is wrong. Critically, large offsets ($>10$~km) are treated as
self-evident correctness violations independent of external ground
truth: no residential household geocoding result with a 56~km or 107~km
round-trip offset can be operationally correct, regardless of what the
geocoder's match-precision field reports.

\subsection{Formal Specification}

Algorithm~\ref{alg:rto} specifies the oracle formally. Parameters are
$\theta_1 = 50$~m (high-accuracy), $\theta_2 = 200$~m (degraded),
$\theta_3 = 2000$~m (false-confidence boundary).

The \Call{Deduplicate}{} step consolidates transactions recorded at the
same coordinate into a single test case. These co-located transactions
are not repeat visits: the deployment was single-pass, and no household
was served twice. Co-location arises structurally, from multi-family
dwellings, adjacent households served at a shared access pin,
vertically stacked residences, and opportunistic pickup at the delivery
vehicle. Consolidation is therefore a property of the addressing
environment rather than an artifact of collection, and it is required
for oracle evaluation because a geocoder queried twice with the same
coordinate returns the same result: leaving co-located transactions in
place would weight the corpus by household density rather than by
distinct locations under test.

\begin{algorithm}[t]
\caption{Round-Trip Offset Oracle}
\label{alg:rto}
\begin{algorithmic}[1]
\Require $C = \{c_1, c_2, \ldots, c_n\}$ (GPS coordinate corpus)
\Require $G$ (geocoding API: $\mathit{reverse}(c) \to a$, $\mathit{forward}(a) \to c'$)
\Require $\theta_1, \theta_2, \theta_3$ (classification thresholds)
\Ensure $R = \{(c_i, \mathit{offset}_i, \mathit{tier}_i)\}$
\Procedure{RoundTripOracle}{$C, G, \theta_1, \theta_2, \theta_3$}
  \State $C' \gets \Call{Deduplicate}{C}$ \Comment{consolidate co-located transactions}
  \For{each $c_i \in C'$}
    \State $a_i \gets G.\mathit{reverse}(c_i)$ \Comment{reverse geocode}
    \If{$a_i = \textsc{NoResult}$}
      \State $\Call{Record}{c_i, \infty, \textsc{NoCoverage}}$; \textbf{continue}
    \EndIf
    \State $c'_i \gets G.\mathit{forward}(a_i)$ \Comment{forward geocode}
    \State $\mathit{offset}_i \gets \Call{Haversine}{c_i, c'_i}$ \Comment{meters}
    \State $\mathit{tier}_i \gets \Call{Classify}{\mathit{offset}_i, \theta_1, \theta_2, \theta_3}$
    \State $\Call{Record}{c_i, \mathit{offset}_i, \mathit{tier}_i}$
  \EndFor
  \State \Return $R$
\EndProcedure
\Function{Classify}{$\delta, \theta_1, \theta_2, \theta_3$}
  \If{$\delta \leq \theta_1$} \Return \textsc{HighAccuracy}
  \ElsIf{$\delta \leq \theta_2$} \Return \textsc{Degraded}
  \ElsIf{$\delta \leq \theta_3$} \Return \textsc{Critical}
  \Else{} \Return \textsc{SilentFailure}
  \EndIf
\EndFunction
\end{algorithmic}
\end{algorithm}

\medskip
\noindent\textbf{Metamorphic relation:} For valid $c_i$,
\textsc{RoundTripOracle} should satisfy $d(c_i,
\mathit{forward}(\mathit{reverse}(c_i))) \leq \theta_1$. Any violation
is an automatically detected correctness failure requiring no human
oracle.

\subsection{Comparison with Alternative Detection Methods}

Table~\ref{tab:detection} provides a structured comparison of four
detection methods across the failure classes documented in this study.
The comparison makes the oracle's distinguishing property explicit:
resolution rate is blind to any failure that returns a result.
Match-precision labels are uncalibrated for out-of-distribution
jurisdictions. Multi-geocoder agreement detects paradigm-mismatch
failures but is blind when all geocoders agree on the same wrong
answer. Round-trip offset is the only method that tests the content of
the result against the ground-truth coordinate automatically.

\begin{table*}[t]
\caption{Detection Method Comparison: Which Methods Detect Which Failure Classes}
\label{tab:detection}
\begin{tabular}{lcccc}
\toprule
\textbf{Failure Class} & \textbf{Res.~Rate} & \textbf{Match Label} & \textbf{Multi-Geocoder} & \textbf{Round-Trip Offset} \\
\midrule
Complete invisibility (\noresult)          & \checkmark & n/a        & \checkmark~(all-fail)       & n/a (no result) \\
Paradigm mismatch (wrong addr., resolved)  & $\times$   & $\times$   & \checkmark~(disagree)       & \checkmark~(offset flags) \\
Municipality contamination ($>50$~km)      & $\times$   & $\times$   & Partial                     & \checkmark~(all 20 detected) \\
POI substitution ($>100$~km)               & $\times$   & $\times$   & Partial                     & \checkmark~(all 10 detected) \\
Territorial miscoding (0\% fwd.~success)   & \checkmark & n/a        & \checkmark~(all-fail)       & n/a (no result) \\
Entrance collapse (200~m--5~km)            & $\times$   & $\times$   & Partial                     & \checkmark~(all 6 clusters) \\
\bottomrule
\end{tabular}
\end{table*}

\subsection{Applicability Conditions}

The oracle requires one precondition: the system must have or be able to
collect operational GPS endpoint coordinates. This is increasingly the
default: GPS confirmation is a standard byproduct of mobile delivery,
emergency dispatch, autonomous vehicle routing, and public service
programs. Any system that logs a successful service event at a physical
location is already generating the corpus the oracle requires.

\subsection{Why Round-Trip Offset Is a New Oracle}

Metamorphic testing~\cite{chen1998metamorphic} addresses the oracle
problem by identifying metamorphic relations and using them as test
assertions. Round-trip offset differs from standard applications in
three ways that emerged from the deployment experience. First, the
metamorphic relation is \emph{discovered} from the API's structure, not
designed. Second, the ``input transformation'' is the API's own output
fed back as input, testing internal consistency against an external
spatial specification. Third, the ground-truth specification is
operational data---GPS delivery coordinates generated by normal system
operation, not a constructed test corpus. The oracle converts
operational byproducts into a continuous verification signal without
requiring test design, manual labeling, or controlled experiments,
enabling CI/CD integration. The round-trip offset produced for each
coordinate is a validation event in the sense formalized by
Padilla~\cite{padilla2026trust,padilla2026trustgeo}: a discrete, outcome-based observation
that updates the long-term trust score of a geolocation entity,
complementing instantaneous positional accuracy with longitudinal
reliability evidence. The TrustGeoScore framework~\cite{padilla2026trustgeo}
operationalizes this directly for rural autonomous systems, integrating
round-trip validation outcomes into trust-aware routing decisions.

\section{Background and Related Work}
\label{sec:related}

This section situates the paper across six bodies of literature. We
begin with geocoding accuracy research and explain why it frames the
problem as data quality rather than software reliability. We then draw
on SE testing theory --- oracle problems, metamorphic testing, and
recent REST API testing work --- to establish the verification
framework the oracle builds on. We survey fault taxonomy traditions,
the limited SE literature on location-dependent software, and
confidence calibration in deployed systems. Each subsection closes by
identifying the gap this paper fills relative to that literature. The
final two paragraphs synthesize the gap and state the novelty claim
explicitly.

\medskip
\noindent\textbf{Geocoding accuracy and positional error.}
Geocoding accuracy and positional error have been extensively studied in
the geospatial and GIS literature~\cite{ratcliffe2001,goldberg2007,zandbergen2008}.
These studies characterize how accurately coordinates map to addresses
under various data quality and algorithmic conditions, and establish
that geocoding precision varies substantially across jurisdictions and
address types. However, this line of work evaluates geocoding as a
\emph{data quality} problem rather than a \emph{software reliability}
problem: it does not address how incorrect-but-confident geocoding
outputs propagate through software systems that treat those outputs as
authoritative, nor does it provide mechanisms for detecting such
failures at runtime.

\medskip
\noindent\textbf{Oracle problems and metamorphic testing.}
{\sloppy
Silent failures and oracle problems in software testing provide a more
relevant framework. Barr et al.~\cite{barr2015oracle} define the test
oracle problem as the challenge of determining whether a program's
output is correct without a ground-truth specification. A geocoding API
returning \civic is an implicit oracle claim. Round-trip offset is
structurally related to metamorphic testing~\cite{chen1998metamorphic}:
it exploits a metamorphic relation inherent in geocoding, detects
violations automatically, and requires no human review. Segura
et al.~\cite{segura2016survey} survey metamorphic testing across domains
and identify its particular strength for programs whose outputs are
difficult to verify independently---precisely the condition geocoding
presents. Existing metamorphic testing applications, however, operate
on programs under the tester's control with designed input
transformations. This paper shows that the technique extends naturally
to \emph{external infrastructure}: the metamorphic relation is
discovered from the API's semantics, the input transformation is the
API's own output fed back as input, and the ground-truth corpus is
operational data rather than a constructed test suite.
\par}

\medskip
\noindent\textbf{Testing external APIs and black-box services.}
A growing body of SE work addresses automated testing of external APIs
and black-box services. The dominant approach uses structural or
schema-level validation: test generators consume OpenAPI specifications
to produce inputs and verify HTTP status codes, response schemas, and
inter-parameter constraints~\cite{golmohammadi2023survey,kim2022rest}.
This line of work has matured significantly in recent years, with
techniques based on fuzzing, reinforcement learning, and property-based
generation applied to RESTful APIs at scale. However, as
Golmohammadi et al.\ note in their survey of 92 API testing
papers~\cite{golmohammadi2023survey}, existing approaches focus
primarily on availability, schema conformance, and structural
correctness --- they do not address the \emph{semantic correctness} of
API outputs relative to real-world ground truth. A geocoding API that
returns a structurally valid \civic response for every input would pass
all schema-level tests while being wrong by kilometers for 10\% of the
addresses one of the three platforms studied here resolves. The closest
related work is Keeper~\cite{wan2022keeper},
which tests software that uses ML APIs by constructing pseudo-inverse
functions to reverse cognitive tasks and judge output correctness.
Round-trip offset applies the same inversion principle to geocoding:
forward-then-reverse recovers the input if the API is correct. The key
difference is that Keeper operates on the \emph{calling software};
this paper operates on the \emph{API's own spatial consistency}
against an external ground-truth specification --- a distinction that
enables deployment-scale verification without modifying the calling
system.

\medskip
\noindent\textbf{Fault taxonomies in dependable systems.}
Fault taxonomies have a long tradition in SE and dependability
engineering. The Avizienis et al.\ dependability taxonomy~\cite{avizienis2004basic}
distinguishes faults, errors, and failures and classifies faults by
their origin, nature, and persistence. The eight-mode taxonomy
presented in Section~\ref{sec:taxonomy} applies this framework to
geocoding: each mode represents a distinct fault class with a
characteristic origin (training distribution mismatch, database gap,
infrastructure misconfiguration), error manifestation (wrong address,
no result, displaced coordinate), and failure consequence (mis-routing,
failed dispatch, systematic displacement). The taxonomy is the first,
to our knowledge, to characterize geocoding failures in SE terms rather
than as positional measurement errors.

\medskip
\noindent\textbf{Location-dependent software and geospatial systems.}
Location-aware software has been studied extensively in the context of
mobile computing, context-aware systems, and location-based services,
with work addressing accuracy requirements, location privacy, and
indoor/outdoor positioning. However, the geocoding reliability problem
--- specifically, the propagation of wrong-but-confident address
resolution through software systems --- has received little attention in
the SE literature. Existing work treats GPS coordinates as inputs and
geocoding as a solved preprocessing step. The governance dimension of
this gap --- that places without formal addresses require different
infrastructure models entirely, and that community-grounded place
representations may be necessary precursors to formal civic address
assignment --- is examined in complementary work by the
author~\cite{padilla2026reliable}. The architectural implications
extend further: trust-scored geolocation management is needed not only
in rural autonomous systems~\cite{padilla2026trustgeo} but across
IoT-enabled smart city infrastructure more broadly~\cite{padilla2026iot}.
This paper treats geocoding as an unverified external dependency and
applies standard SE verification techniques to it, exposing a class of
failures that is invisible to the evaluation methodology the geospatial
literature relies on.

\medskip
\noindent\textbf{Confidence calibration in deployed systems.}
{\sloppy
The overconfidence documented in this study---Google results labeled
\loctype = \rooftop correct by round-trip criteria only 50.6\% of the
time---is analogous to the
calibration failures documented in production ML
systems~\cite{guo2017calibration}, where confidence scores
systematically fail on out-of-distribution inputs. Calibration research
in ML has produced practical techniques including temperature
scaling~\cite{guo2017calibration} and conformal prediction for
quantifying and correcting overconfidence at deployment time. No
analogous calibration framework exists for geocoding APIs, which expose
categorical match-precision labels but provide no mechanism for
distribution shift detection or recalibration. Round-trip offset
provides a
practical substitute: it replaces the geocoder's stated match precision
with a measured spatial accuracy signal that is empirically grounded in
the deployment distribution rather than the geocoder's training
distribution.
\par}

\medskip
\noindent\textbf{The gap this paper fills.}
The works surveyed above share a common assumption: the component under
test is either under the developer's control or can be fully specified.
Property-based testing constructs inputs with known correct outputs.
Metamorphic testing designs relations for programs the tester can
instrument. API testing frameworks require OpenAPI specifications to
generate structurally valid inputs. None of these approaches addresses
the case where a third-party API makes \emph{implicit semantic
claims} --- correctness assertions embedded in output fields such as
result type and match-precision label --- that the calling system cannot
verify without an external ground-truth oracle. This paper fills that
gap. Rather than testing a program under the developer's control, we
treat external geolocation infrastructure as an unverified component
and derive a correctness oracle directly from its semantics and
operational data. To our knowledge, this is the first
deployment-scale demonstration of a metamorphic oracle for verifying
the \emph{semantic correctness} of a third-party location API in a
production system, using operational GPS data as the ground-truth
specification without requiring manual labeling, constructed test
inputs, or access to the API internals.

\medskip
\noindent\textbf{Synthesis.}
No prior work connects these threads: the geospatial literature treats
positional error as a data quality concern; the SE testing literature
frames oracle failures as a verification concern; the ML literature
frames overconfidence as a calibration concern. This paper connects all
three by showing that round-trip offset exposes failures that
simultaneously manifest as positional error, oracle failure, and
confidence miscalibration, and that it can be deployed as a standard
automated test within existing SE workflows.

\section{Evaluation Setup}
\label{sec:setup}

We evaluate the oracle against a deployment-scale corpus of 24,501
unique operationally validated GPS coordinates, consolidated from
32,019 delivery transactions in a house-to-house food
assistance program in Bayam\'on, Puerto Rico. This section describes the
deployment context and data quality, the oracle methodology as applied
to this dataset, reproducibility provisions, and the ground truth
validation methodology. The deployment context is the evaluation
instrument, not the subject: it provides an unusually strong ground
truth against which oracle behavior can be measured at scale.

\subsection{Deployment Context}

This study analyzes 32,019 GPS delivery transactions from a single USDA
COVID-19 food assistance deployment in Bayam\'on, Puerto Rico,
conducted during the 2020
pandemic. The broader program enrolled approximately 35,000 households;
this study uses the 32,019 GPS-pin records for which validated delivery
coordinates were recorded. Each delivery pin was recorded at the point
of successful delivery and validated for use in official municipal food
delivery reports.

The 32,019 transactions resolve to 24,501 unique coordinates. The
deployment was single-pass: no household was served twice, and the
difference between the two figures is therefore not repeat delivery.
Co-location arises from the structure of the housing stock and the
delivery protocol: multi-family dwellings, adjacent households served
at a shared access pin, vertically stacked residences, and
opportunistic pickup at the delivery vehicle by nearby residents under
pandemic movement restrictions. These sources are not separable in the
record. Two consequences follow: 24,501 is a count of unique service
coordinates, not of dwellings and not of households, and households
served at a pickup point rather than at their door are not represented
by a coordinate of their own. All oracle evaluation in Section~\ref{sec:results} is
performed over the 24,501 unique coordinates, since a geocoder returns
an identical result for repeated queries at the same coordinate and
retaining co-located transactions would weight the corpus by household
density rather than by distinct locations under test. Household counts
exceeding pin counts in Table~\ref{tab:centroid_geocoder} are a direct
consequence of this same co-location structure.

A 100-pin pilot run (\texttt{random\_state=42})
revealed a 52\% Google--Nominatim disagreement rate, prompting the
addition of HERE as a third geocoder~\cite{google2024,nominatim2024,here2024}.\footnote{Google
offers a separate Address Validation API with dedicated handling for
Puerto Rico urbanization names~\cite{googleav}. That product is outside
the scope of this evaluation, which targets the standard Geocoding
API: the integration path available to a location-dependent system
without jurisdiction-specific configuration. The existence of a
separate specialized product is itself consistent with the failures
documented here being structural rather than incidental.}
The same geolocation corpus is used in complementary work to evaluate
spatial structure, infrastructure proximity, and namespace scalability
for trust-aware autonomous routing~\cite{padilla2026trustgeo}.

Dataset parameters are summarized in Table~\ref{tab:dataset}. A
pre-analysis audit confirmed exceptional data quality: zero missing,
zero-value, or out-of-bounds coordinates; zero low-precision entries.
56\% of pins were recorded at eight decimal places (sub-centimeter
theoretical precision); 319 pins (1.0\%) fell in adjacent
municipalities.

\begin{table}[t]
\caption{Dataset Parameters}
\label{tab:dataset}
\begin{tabular}{ll}
\toprule
\textbf{Parameter} & \textbf{Value} \\
\midrule
Program enrollment         & $\approx$35,000 households \\
Total delivery transactions & 32,019 \\
Unique GPS coordinates     & 24,501 (co-location, not revisits) \\
Municipalities covered     & Bayam\'on + 2--3 adjacent \\
Coverage area              & 199.7~km$^2$ centered on Bayam\'on \\
Deployment period          & 2020 (COVID-19 pandemic) \\
Geocoding period           & 2025 \\
Delivery model             & House-to-house, single pass \\
Data format                & CSV: geoid, geolong, geolat \\
PII in dataset             & None (opaque identifiers only) \\
IRB status                 & Exempt (no PII, opaque coordinates) \\
Federal validation         & Federal program delivery records \\
\bottomrule
\end{tabular}
\end{table}

No names, household composition, or individual-level data
appear anywhere in this paper or its released artifacts. Case study
examples use structural descriptions rather than specific coordinates or
property identifiers; the deployment dataset is not released
(see Section~\ref{sec:das}).

\subsection{Round-Trip Offset Methodology}

The oracle design and formal specification are presented in
Section~\ref{sec:oracle}; Algorithm~\ref{alg:rto} shows the
implementation as applied to this dataset. Each geocoder was evaluated
using its own round-trip (Google reverse $\to$ Google forward, etc.) to
measure internal spatial consistency against the ground-truth delivery
pins. Round-trip measurement detects false-confidence failures that
resolution-rate evaluation cannot: it tests output content, not just
output presence.

{\sloppy
Google was geocoded twice. The initial run (Run~1) completed
successfully but was found to have omitted Google's \loctype
field --- which reports whether a result is \rooftop,
\texttt{RANGE\_INTERPOLATED}, \texttt{GEOMETRIC\_CENTER}, or
\texttt{APPROXIMATE} --- from the stored output. A complete rerun
(Run~2) was performed with the script corrected to capture the full
field set. Resolution rates between Run~1 and Run~2 were negligibly
different, confirming that the omission did not affect the
classification results. All Google figures reported in this paper use
Run~2, which contains the complete field set including
\loctype.
\par}

\subsection{Reproducibility}

The round-trip offset oracle pipeline is released as a self-contained
package with a synthetic 500-coordinate corpus (see
Section~\ref{sec:das}). The deployment dataset is not released; see
Section~\ref{sec:das} for the full explanation.

\subsection{Ground Truth Validation}

Ground truth validation was conducted in two phases. In the first phase,
a stratified manual validation sample of 58 cases was drawn from the
pilot geocoding run, covering all agreement and disagreement categories
between Google and Nominatim: both-fail ($n=2$), Nominatim-wins
($n=35$), Google-wins ($n=6$), and both-agree-CIVIC ($n=15$). Each case
was validated via Street View, OSM imagery, and direct local knowledge.
In the second phase, a subset was cross-referenced against Puerto Rico's
CRIM property registry~\cite{crim} by coordinate-based lookup, independent of any
geocoder output, confirming that Nominatim's urbanization lot format
aligns with official records while Google's sequential street-number
format does not, and that two BOTH\_FAIL locations are officially
registered CRIM properties. The validation sample is not released; it
contains coordinate-to-property linkages that carry the same sensitivity
as the deployment dataset.

HERE was added after the pilot run and so did not contribute to the
stratification of the 58-case sample, though it was subsequently run
against those cases. Its results are validated through two
complementary mechanisms: (1)~multi-geocoder agreement with Nominatim
on the 11,262 coordinates where all three geocoders return
\civic, providing cross-platform confirmation for the intersection
set; and (2)~round-trip offset distribution analysis, in which HERE's
91.2\% sub-50m rate and structurally separate catastrophic-failure tail
are consistent with the bimodal pattern confirmed across all three
geocoders. No cases were found in manual review where HERE produced a
plausible-looking result that Nominatim or CRIM contradicted.

\section{Detection Results and Oracle Validation}
\label{sec:results}

This section reports oracle validation results across three geocoders
and 24,501 unique coordinates. We first establish the resolution rate
baseline, then analyze three-way agreement patterns and a cross-cutting
infrastructure failure, then present round-trip accuracy results and the
bimodal distribution structure that makes the oracle work, and finally
draw out the implications for autonomous systems. The central finding is
that standard resolution-rate evaluation and round-trip offset select
for opposite behaviors: the cases resolution rate counts as successes
are the cases round-trip offset identifies as dangerous.

\subsection{Overall Resolution Rates}

Table~\ref{tab:resolution} reports geocoding resolution rates across all
24,501 unique coordinates for three geocoders. The ``Any geocoder
(union)'' Unresolved row (2,441 coordinates, 10.0\% of the 24,501
unique coordinates) represents
locations for which no geocoder returned a \civic address---the
irreducible invisible population that cannot be resolved regardless of
geocoder selection. This 10.0\% figure is a property of the corpus, not
of any individual platform, and should not be conflated with the
separate, Google-specific 10.0\% reported in
Section~\ref{sec:roundtrip}, whose denominator is
Google's 13,221 resolved addresses.

\begin{table*}[t]
\caption{Overall Geocoding Resolution Rates ($N = 24{,}501$ unique coordinates)}
\label{tab:resolution}
\begin{tabular}{lrrrl}
\toprule
\textbf{Geocoder} & \textbf{CIVIC\_ADDRESS} & \textbf{ROAD\_ONLY} & \textbf{Unresolved} & \textbf{Notes} \\
\midrule
Google                & 13,221 (54.0\%) & 4,883 (19.9\%) & 6,397 (26.1\%) & Rerun to capture \loctype; rates unchanged \\
Nominatim             & 18,719 (76.4\%) & 4,039 (16.5\%) & 1,743 (7.1\%)  & \texttt{countrycodes=US} fix \\
HERE (corrected)      & 19,727 (80.5\%) & 4,319 (17.6\%) & 455 (1.9\%)    & \resulttype classifier \\
HERE (uncorrected)    & 20,419 (83.3\%) & ---            & ---            & Before correction \\
Any geocoder (union)  & 22,060 (90.0\%) & ---            & 2,441 (10.0\%) & No geocoder returned \civic \\
All three agree       & 11,262 (46.0\%) & ---            & ---            & Intersection \\
\bottomrule
\end{tabular}

\footnotesize The three-way agreement analysis in
Table~\ref{tab:agreement} is computed over HERE's uncorrected
classification (20,419); the all-fail count of 2,441 is therefore a
lower bound under the corrected classifier.
\end{table*}

\subsection{Three-Way Agreement Analysis}

Table~\ref{tab:agreement} shows three-way geocoder agreement patterns.
6,380 locations (26\%) are resolved by both HERE and Nominatim but not
by Google---paradigm mismatch, not a coverage gap. A further 2,441
locations (10.0\%) are invisible to all three geocoders simultaneously:
no geocoder returned a \civic address for these coordinates, meaning
they cannot be resolved by any of the three platforms regardless of
which one a system uses.

Manual validation revealed an important subtlety: two systems can agree
on result \emph{type} while disagreeing on the address \emph{itself}. In
multiple cases, Google returned an incorrect street while
Nominatim correctly identified the urbanization address, despite both
being classified as \civic. This means that category-level agreement
metrics underestimate true geocoder disagreement in informal addressing
contexts. The agreement rates in Table~\ref{tab:agreement} are a
conservative lower bound on actual semantic disagreement---systems that
agree on returning \civic may be returning different, incompatible
addresses for the same coordinate. Round-trip offset detects this class
of disagreement directly; agreement metrics do not.

\begin{table}[t]
\caption{Three-Way Geocoder Agreement Patterns}
\label{tab:agreement}
\begin{tabular}{lrrl}
\toprule
\textbf{Pattern} & \textbf{Count} & \textbf{\%} & \textbf{Implication} \\
\midrule
All three CIVIC\_ADDRESS & 11,262 & 46.0 & Highest confidence \\
Only Google fails        &  6,380 & 26.0 & Paradigm mismatch \\
Only Nominatim fails     &    953 &  3.9 & OSM coverage gap \\
Only HERE fails          &    442 &  1.8 & HERE coverage gap \\
All three fail           &  2,441 & 10.0 & Structurally invisible \\
\bottomrule
\end{tabular}
\end{table}

\subsection{The \texttt{countrycodes} Configuration Failure}

{\sloppy
Standard practice of passing \texttt{countrycodes=PR} to Nominatim
returns zero results because Puerto Rico is indexed under
\texttt{countrycodes=US}. This produced 100\% forward geocoding failure
across 18,719 Nominatim queries---a visible, not silent, failure that
resolution-rate evaluation correctly detects. However, it is a
\emph{developer trap}: the failure arises from following standard API
documentation, not from a geocoding error. Correcting to
\texttt{countrycodes=US} restored 99.8\% forward success. This
illustrates how territorial status creates systematic configuration
failures invisible to developers but detectable by resolution rate.
\par}

\subsection{Round-Trip Accuracy Results}
\label{sec:roundtrip}

Table~\ref{tab:roundtrip} reports round-trip offset results. Nominatim
achieves the lowest median offset (12.2~m) and zero catastrophic
failures. HERE achieves the highest resolution rate (80.5\%, median
13.6~m). Google achieves the lowest resolution rate (54.0\%, median
49.2~m). The false-confidence rate---the share of a platform's own
resolved addresses with round-trip offsets above 2~km---is 10.0\% for
Google (1,322 of its 13,221 resolved coordinates, or 5.4\% of the
24,501-coordinate corpus), against 0.5\% for HERE and 0.04\% for
Nominatim. Google's rate is thus 250 times Nominatim's and 20 times
HERE's; the effect is specific to one platform rather than
characteristic of commercial geocoders as a class. We report the platform's result classification rather than a scalar
score because the geocoding APIs evaluated here expose result-type
categories rather than a single confidence value: \rooftop is the
highest-precision location classification Google's interface exposes,
and \housenum is HERE's.

Nominatim's strong performance is partly attributable
to the OpenStreetMap disaster mapping that followed Hurricane
Mar\'ia in 2017, when volunteers traced roads, buildings, and access
points across Puerto Rico that no commercial database had
captured~\cite{hot2017,osm2024}. This should not be read as a
recommendation to change vendors: it is not an algorithmic result but
an instance of the mechanism this paper argues for, local place
knowledge captured once and made available to the systems that
follow.

\begin{table}[t]
\caption{Round-Trip Offset Results by Geocoder}
\label{tab:roundtrip}
\begin{tabular}{lrrr}
\toprule
\textbf{Metric} & \textbf{Google} & \textbf{HERE} & \textbf{Nominatim} \\
\midrule
CIVIC\_ADDRESS count    & 13,221   & 19,727    & 18,719  \\
Forward success rate    & 99.8\%   & 99.2\%    & 99.8\%  \\
Median offset           & 49.2~m   & 13.6~m    & 12.2~m  \\
Mean offset             & 554.7~m  & 67.3~m    & 22.4~m  \\
Max offset              & 56,145~m & 107,008~m & 2,286~m \\
$<50$~m (high)          & 50.6\%   & 91.2\%    & 95.3\%  \\
50--200~m               & 29.1\%   &  4.3\%    &  4.0\%  \\
200--500~m              &  3.4\%   &  1.4\%    &  0.2\%  \\
500~m--2~km             &  6.9\%   &  2.6\%    &  0.4\%  \\
$>2$~km (false conf.)   & 10.0\%   &  0.5\%    & 0.04\%  \\
Catastrophic $>10$~km   & 20 cases & 10 cases  & 0 cases \\
\bottomrule
\end{tabular}

\footnotesize All percentages are computed over each platform's own
\civic count, not over the 24,501-coordinate corpus.
\end{table}

\subsection{Detection Performance: Bimodal Distribution}

Table~\ref{tab:roundtrip} reveals a bimodal offset distribution that
explains why round-trip offset works as a detector. The confirmed-correct
population ($<50$~m) and confirmed-wrong tail ($>2$~km) are structurally
separated: catastrophic failure modes (municipality contamination, POI
substitution) produce offsets of 56--107~km, not 2.1~km. A threshold
anywhere between 50~m and 2~km separates the two populations without
overlap. Google's distribution differs notably: only 50.6\% fall under
50~m versus 91.2\% (HERE) and 95.3\% (Nominatim). The 50~m--2~km
ambiguous zone---5,199 Google cases (39.4\%), 1,681 HERE (8.3\%), 872
Nominatim (4.7\%)---represents the honest operational boundary of the
methodology, requiring additional validation before autonomous action.

\subsection{Implications for Autonomous Systems}

The most dangerous failures---large offsets indicating catastrophically
wrong addresses---occur under each platform's strongest match-precision
label with no error signal.
A \noresult response is operationally safe: the system retains the GPS
coordinate and can route directly to it. A \civic response with a large
round-trip offset is operationally dangerous: the system routes to the
wrong location with no signal that anything is wrong. Resolution rate
penalizes the safe failure and ignores the dangerous one.

The contrast is concrete. In one confirmed case (entrance collapse,
Failure Mode~3), Google returns \noresult---a clear signal that the
location cannot be resolved autonomously. The system escalates to human
review. The failure is acknowledged and safe. In a second confirmed
case, HERE returns \civic with \resulttype set to \housenum---its
highest match precision---and a
1,222~m offset. The system has no signal that anything is wrong. It
dispatches autonomously to a highway entrance with 1.2~km of private
road remaining. The household receives nothing. The system logs a
successful geocoding event. The first failure is visible; the second is
silent. Standard evaluation rewards the second and penalizes the first.

\section{Fault Taxonomy}
\label{sec:taxonomy}

Table~\ref{tab:taxonomy} provides a structured overview of eight
distinct failure modes. The modes partition evenly into two classes.
Four are \emph{software-solvable}: paradigm mismatch, municipality
contamination, POI substitution, and address hallucination each arise
from how a platform models addresses or from gaps in its coverage, and
a better geocoder, a validation layer, or an automated oracle can
detect and correct them. Four are not. Complete invisibility and
territorial status miscoding follow from the absence or
misclassification of civic addressing infrastructure; entrance collapse
and property centroid displacement follow from rural addressing
conventions that anchor civic identity to highway access rather than to
dwellings. No software improvement resolves an address that was never
assigned, or recovers a dwelling location that no address string
specifies.

\begin{table*}[t]
\caption{Taxonomy of Silent Failures in Location-Dependent Software}
\label{tab:taxonomy}
\begin{tabular}{lllrl}
\toprule
\textbf{Failure Mode} & \textbf{Geocoder} & \textbf{Detection} & \textbf{Max Offset} & \textbf{Consequence} \\
\midrule
Paradigm mismatch          & Google   & CRIM comparison            & $\sim$200~m      & Wrong street in urbanization \\
Complete invisibility      & All three & \noresult flag            & n/a              & Location not in database \\
Entrance collapse          & HERE     & Round-trip + cluster       & 233~m--5.1~km    & Anchors to highway, not dwelling \\
Property centroid displ.   & All three & Round-trip + addr.~type   & 100~m--2~km      & Dwelling displaced inside parcel \\
Municipality contamination & Google   & Round-trip $>10$~km        & 56~km            & Routes to different city \\
POI substitution           & HERE     & Round-trip $>10$~km        & 107~km           & Routes to business, not home \\
Address hallucination      & Nominatim & Round-trip + partial id   & $\sim$100~m      & Borrows nearby address fragment \\
Territorial status miscoding & Nominatim & Fwd.~geocoding failure  & 100\% fwd.~fail  & \texttt{countrycodes=PR} returns zero \\
\bottomrule
\end{tabular}

\footnotesize The Max Offset column reports observed maxima except for
entrance collapse, where the range gives per-cluster medians
(Table~\ref{tab:clusters}).
\end{table*}

\medskip
\noindent\textbf{Failure Mode 1---Paradigm Mismatch.} Google Maps
applies continental US sequential street numbering to Puerto Rico's
urbanization lot-code system. The two paradigms are structurally
incompatible: a geocoder trained on ``235 Calle Caracas'' will fail
systematically on ``[lot-code], [street], [urbanization name].'' A
CRIM-registered urbanization property: Google returns a sequential
street number on the wrong street (87~m offset); HERE returns the
correct urbanization lot identifier (22~m); Nominatim returns the
correct street. The failure is paradigm-specific, not a database
coverage gap.

\medskip
\noindent\textbf{Failure Mode 2---Complete Invisibility.} 2,441
locations (10.0\% of the 24,501 unique coordinates) could not be
resolved by any of the three geocoders. Two of these were confirmed
against the CRIM registry as officially registered properties: the
coordinate names a real, legally recorded parcel that no platform can
express as a civic address. Invisibility here is a property of the
addressing infrastructure rather than of the location.

\medskip
\noindent\textbf{Failure Mode 3---Entrance Collapse.} Entrance collapse
occurs when a geocoder resolves a rural property coordinate to the
nearest public highway access point. Table~\ref{tab:clusters} shows six
confirmed clusters identified by address-level cluster analysis.

\begin{table}[t]
\caption{Entrance Collapse Clusters}
\label{tab:clusters}
\begin{tabular}{lrrr}
\toprule
\textbf{Address} & \textbf{Coords} & \textbf{GPS Spread} & \textbf{Median Offset} \\
\midrule
KM 5.4 PR-830  & 11 & 168~m & 1,149~m \\
KM 9.4 PR-829  &  4 & 171~m &   596~m \\
KM 3.3 PR-862  &  4 & 150~m &   233~m \\
KM 12.4 PR-167 &  4 & 113~m & 5,133~m \\
KM 8.3 PR-812  &  2 & 260~m &   378~m \\
KM 0.3 PR-871  &  2 & 113~m &   397~m \\
\bottomrule
\end{tabular}
\end{table}

\medskip
\noindent\textbf{Failure Mode 4---Property Centroid Displacement.}
In rural areas, civic addresses anchor to highway reference points
rather than dwelling locations. A geocoder resolving the address
correctly---placing the result at the KM marker---is technically
accurate but operationally incorrect. The dwelling is located somewhere
inside the property, often hundreds of meters to kilometers from the
highway anchor, at a location no address string specifies. All three
geocoders exhibit this through different mechanisms: HERE resolves
kilometer marker addresses to highway entrance points (280 pins, 357
households); Nominatim resolves carretera street numbers to the highway
alignment rather than the dwelling set back from the road (180 pins,
227 households); Google resolves highway references to the nearest road
segment (124 pins, 148 households). Household counts exceed pin counts
because of the co-location structure described in
Section~\ref{sec:setup}. Tables~\ref{tab:centroid_dist}
and~\ref{tab:centroid_geocoder} show the offset distribution and
geocoder breakdown. Centroid displacement is not a geocoder-specific
quirk but a structural consequence of rural addressing systems that
anchor civic identity to highway access rather than dwelling location.

\begin{table}[t]
\caption{Property Centroid Displacement: Offset Distribution ($N = 584$)}
\label{tab:centroid_dist}
\setlength{\tabcolsep}{4pt}
\begin{tabular}{lrp{3.2cm}}
\toprule
\textbf{Offset Range} & \textbf{Pins} & \textbf{Typical Context} \\
\midrule
100--200~m   & 220 (37.7\%) & Small parcel / near-highway \\
200--300~m   &  98 (16.8\%) & Mid-size rural property \\
300--500~m   &  51 (8.7\%)  & Large parcel, set back \\
500~m--1~km  &  99 (17.0\%) & Large parcel / gated access \\
1--2~km      & 116 (19.9\%) & Farm or remote dwelling \\
\bottomrule
\end{tabular}
\end{table}

\begin{table}[t]
\caption{Property Centroid Displacement by Geocoder}
\label{tab:centroid_geocoder}
\setlength{\tabcolsep}{4pt}
\begin{tabular}{lrrp{2.4cm}}
\toprule
\textbf{Geocoder} & \textbf{Pins} & \textbf{HH} & \textbf{Mechanism} \\
\midrule
HERE      & 280 & 357 & KM marker entrance \\
Nominatim & 180 & 227 & Carretera at highway \\
Google    & 124 & 148 & Highway ref.\ point \\
\midrule
\textbf{Total} & \textbf{584} & \textbf{732} & All three affected \\
\bottomrule
\end{tabular}
\end{table}

\medskip
\noindent\textbf{Failure Mode 5---Municipality Contamination.}
In 20 Google cases, incorrect municipality names are embedded in the
returned address, producing offsets up to 56~km. The geocoder resolves
the address to a real location --- but in the
wrong city. The calling system receives a \civic result labeled
\loctype = \rooftop, Google's strongest match-precision value, and no
indication that the returned
coordinate is 56 kilometers from the delivery pin. Detection requires
round-trip offset; neither resolution rate nor the match-precision
label provides any signal.

\medskip
\noindent\textbf{Failure Mode 6---POI Substitution.}
In 10 HERE cases, a nearby point of interest --- a business, landmark,
or public facility --- is substituted for the residential address,
producing offsets up to 107~km. The substitution is structurally
coherent: the returned address is a real, geocodable place. It is
simply not the household. This is the failure mode with the largest
observed offset in the dataset (107~km), returned as \civic under
HERE's highest-precision \resulttype classification.

\medskip
\noindent\textbf{Failure Mode 7---Address Hallucination.}
Nominatim borrows partial address identifiers from nearby properties
in sparse rural areas where the target address has incomplete OSM
coverage. The returned address is plausible in structure but
corresponds to a neighboring lot, producing offsets typically under
100~m. Detectable by partial-identifier consistency check against
CRIM; not detectable by round-trip offset alone at the sub-100m
scale.

\medskip
\noindent\textbf{Failure Mode 8---Territorial Status Miscoding.}
{\sloppy Standard geocoding practice recommends specifying a country
code to improve result accuracy (e.g., \texttt{countrycodes=PR} for
Puerto Rico). Because Puerto Rico is indexed as a US territory rather
than an independent country, passing \texttt{countrycodes=PR} returns
zero results---a 100\% forward geocoding failure affecting all 18,719
Nominatim \civic results in this study. The fix
(\texttt{countrycodes=US}) is non-obvious and undocumented in
Nominatim's primary API reference. This failure is cross-cutting: it
affects any developer following standard API documentation for Puerto
Rico, and generalizes to all US territories and any jurisdiction whose
country code status differs from developer expectations. It is also the
one mode in the second class that a single configuration change
resolves; it is grouped there because the trap originates in how
territorial status is encoded rather than in the calling code.\par}

\smallskip
\noindent\fbox{\parbox{0.95\columnwidth}{\small\textbf{Critical Finding:}
Every large-offset case---all 30 instances exceeding 10~km---was
returned as a \civic result carrying its platform's strongest
match-precision label---\loctype = \rooftop for Google's 20 cases, and
HERE's highest-precision \resulttype classification for its 10---and no
error field. Offsets of
this magnitude are operationally implausible as correct resolutions and
are treated as self-evident correctness violations. Round-trip offset
detected all 30; resolution-rate evaluation detected zero.}}

\section{Discussion}
\label{sec:discussion}

This section interprets the empirical findings from three perspectives.
First, we examine how failure modes distribute across urban and rural
zones, revealing that the same deployment contains two structurally
different failure regimes requiring different engineering responses.
Second, we analyze the false confidence problem and its implications for
CI/CD integration. Third, we state the study's limitations and distill
six concrete, automatable lessons for practitioners building
location-dependent systems.

\subsection{Spatial Context and Two Failure Regimes}

Tables~\ref{tab:composition} and~\ref{tab:spatial} reveal two
structurally different failure regimes.

\begin{table}[t]
\caption{Dataset Composition by Area Type}
\label{tab:composition}
\begin{tabular}{lrr}
\toprule
\textbf{Area Type} & \textbf{Coords} & \textbf{\%} \\
\midrule
Urban (urbanization)  &  8,904 & 36.3 \\
Rural (km/sector)     & 13,569 & 55.4 \\
Unknown               &  2,028 &  8.3 \\
\bottomrule
\end{tabular}
\end{table}

\begin{table}[t]
\caption{Geocoding Resolution Rates by Area Type}
\label{tab:spatial}
\begin{tabular}{lcc}
\toprule
\textbf{Metric} & \textbf{Urban} & \textbf{Rural} \\
\midrule
Google \civic      & 66.0\% & 46.2\% \\
Nominatim \civic   & 93.3\% & 67.8\% \\
HERE \civic        & 94.1\% & 71.9\% \\
All three fail     &  1.4\% & 18.4\% \\
Only Google fails  & 28.9\% & 23.8\% \\
\bottomrule
\end{tabular}
\end{table}

\medskip
\noindent\textbf{Regime A---Formal Addressing (Urban).} 94.1\% of urban
coordinates are resolved by HERE and 93.3\% by Nominatim; only 1.4\%
are invisible to all three. The dominant failure is paradigm mismatch:
switching from Google to HERE raises resolution from 66.0\% to
94.1\%---a 28-percentage-point improvement---without any change to
the underlying address data. Urban geocoding failure is almost entirely
a software-layer problem.

{\sloppy
\medskip
\noindent\textbf{Regime B---Informal Addressing (Rural).} In the rural
zones, 18.4\% of coordinates are invisible to all three geocoders
simultaneously---a 13-fold higher all-fail rate than urban areas.
Switching geocoders raises rural resolution from 46.2\% to 71.9\%, but
18.4\% remain invisible regardless of geocoder choice. Rural geocoding
failure is not a software-layer problem: it requires formal address
assignment, community mapping, and operational knowledge preservation~\cite{census2022opportunity,padilla2026reliable}.
The 13-fold urban/rural gap is not specific to this municipality: any
deployment spanning formal and informal addressing zones will encounter
the same structural divide.
\par}

\subsection{The False Confidence Problem}

The false confidence problem is a broader automated SE failure class:
systems that consume uncalibrated match-precision signals from external
components. Google results labeled \loctype = \rooftop are correct by
round-trip criteria only 50.6\% of the
time---miscalibrated by any standard~\cite{guo2017calibration}.
Table~\ref{tab:thresholds} maps offset ranges to CI/CD action gates.
A pipeline that fails when median offset exceeds 50~m, or when any
result exceeds 2~km, will catch geocoder degradation automatically.
These tiers correspond directly to the trust-based routing gates in
TrustGeoScore~\cite{padilla2026trustgeo}, which integrates equivalent
offset thresholds into a Dijkstra-based routing extension for rural
autonomous systems, and to the trust threshold API exposed by the
IoT smart city architecture~\cite{padilla2026iot}.

\begin{table}[t]
\caption{Automated Threshold Selection for CI/CD Integration}
\label{tab:thresholds}
\begin{tabular}{lll}
\toprule
\textbf{Round-Trip Offset} & \textbf{Accuracy Tier} & \textbf{CI/CD Gate} \\
\midrule
$< 50$~m    & High           & Act autonomously \\
50--200~m   & Degraded       & Verify before acting \\
200--500~m  & Low            & Human confirmation \\
500~m--2~km & Critical       & Do not act autonomously \\
$> 2$~km    & Silent Failure & Alert + human review \\
\noresult   & No Coverage    & Cannot initialize \\
\bottomrule
\end{tabular}

\footnotesize The 200--500~m and 500~m--2~km gates subdivide
Algorithm~\ref{alg:rto}'s \textsc{Critical} tier; the subdivision is an
operational refinement, not a separate threshold parameter.
\end{table}

Any system gating autonomous action on geocoding results should treat
the geocoder's match-precision label as an uncalibrated prior.
Round-trip offset replaces
it with a measured spatial accuracy signal: a 12~m offset warrants high
confidence; a 5,133~m offset warrants zero autonomous action, regardless
of whether the result is labeled \housenum or \rooftop.

\subsection{Limitations}

This study has seven limitations. (1)~All coordinates are from a single
municipality and a single collection context; failure rates will differ
elsewhere, though the eight failure modes generalize as a class.
(2)~Delivery events are from 2020 and were geocoded in 2025; HERE and
Nominatim databases have been updated since collection.
Google's paradigm mismatch is structural and unlikely to have
changed materially. (3)~The \civic classification schema is
operationally motivated but not universal; the released pipeline supports
replication with alternative definitions. (4)~The HERE classifier
correction uses HERE's own \resulttype field; uncorrected rates
inflate \civic counts by 692 coordinates (2.8 percentage points), with
both reported in
Table~\ref{tab:resolution}. (5)~The 58-case CRIM ground truth validation
is confirmatory, not exhaustive---cases were selected to represent
diverse failure modes rather than a random sample. The validation sample
is not released; it contains coordinate-to-property linkages carrying
the same sensitivity as the deployment dataset. (6)~The corpus contains
no record of operational failure: a coordinate exists only where a
delivery succeeded, so every one of the 24,501 locations was reachable
and no operational failure rate can be computed from these data. The
failures reported here are failures of address-to-coordinate
translation, not of reachability. (7)~The coordinates are the records
submitted to federal agencies and had passed the operator's own
validation before submission; readings rejected during that validation
are not present, so the corpus is the accepted record rather than the
complete field capture. The deployment is also a single wave, so no
decay rate for operational validity can be measured from it.

\subsection{Lessons Learned for Automated SE Practice}

The deployment experience yields six concrete, automatable lessons:

\begin{enumerate}
\item \textbf{Measure outputs, not availability.} Resolution rate is
  necessary but not sufficient. Round-trip offset is the correctness
  criterion that matters for autonomous action.

\item \textbf{Treat geocoder match-precision labels as uncalibrated.}
  Fields such as \loctype and \resulttype are categorical match
  descriptors, not calibrated probabilities, and they are not
  calibrated for jurisdictions outside their
  training distribution. Do not gate autonomous action on them alone.

\item \textbf{Automate round-trip offset in CI/CD pipelines.}
  Integrate round-trip offset as an automated regression test. API
  access and a GPS corpus are the only requirements; geocoder degradation
  is caught automatically without \mbox{human} review.

\item \textbf{Use multi-geocoder agreement as a trust signal.}
  A result confirmed by two or more geocoders at small round-trip offset
  warrants higher trust. Disagreement is a signal to investigate, not to
  average.

\item \textbf{Preserve operational location knowledge.} GPS pins from
  successful service events are ground-truth location data; retain and
  reuse them across deployments. Treating geolocations as persistent
  entities whose reliability evolves through repeated validation events
  --- rather than as static coordinate tuples --- provides a formal
  basis for this practice~\cite{padilla2026trust,padilla2026trustgeo}.

\item \textbf{Automate detection of silent locale assumptions.}
  Include locale-specific automated test cases that assert against
  zero-result failures and round-trip offset regressions. No human
  oracle is required: the GPS coordinate is the specification.
\end{enumerate}

Two generalizability claims follow. First, the failure class will
manifest in any system using geocoding as a trusted oracle in a
jurisdiction with non-standard addressing. Second, the
software-solvable vs.\ governance-solvable stratification will hold
wherever the formal/informal addressing distinction exists. Specific
rates will differ; the structural distinction will not.

\section{Conclusion}
\label{sec:conclusion}

{\sloppy
This experience paper reports deployment-scale evidence of two distinct
geocoding failure classes with different denominators. Across 32,019
operationally validated delivery transactions in Bayam\'on, Puerto
Rico, consolidated to 24,501 unique coordinates: 2,441 coordinates
(10.0\% of the corpus) are invisible to all three geocoders---an
acknowledged failure class that resolution rate correctly detects, and
a property of the corpus rather than of any one platform; and,
separately, 10.0\% of the 13,221 addresses Google resolves carry
$>2$~km round-trip errors with no error signal---a silent failure
class that resolution rate misses entirely, amounting to 5.4\% of the
corpus and occurring at 250 times the community geocoder's rate and 20
times the second commercial platform's. The silent failure class is
platform-specific rather than a general property of commercial
geocoding. A free community geocoder outperforms both commercial
platforms here, which reflects post-disaster volunteer mapping rather
than algorithmic superiority and is not a vendor recommendation; and
urban and rural failure require different solutions.
\par}

The central lesson: location is a core SE abstraction treated as solved;
it is not. Resolution rate measures availability; round-trip offset
measures correctness. In non-standard addressing environments these
metrics select for opposite behaviors. The failure class is not a
geocoding problem---it is a software engineering problem. The fix is
automated verification, not better geocoders.

The oracle generalizes to any system satisfying the two applicability
conditions in Section~\ref{sec:oracle}. The methodology is the
contribution; Puerto Rico is the stress-test context.

This paper is part of a broader research program addressing geolocation
reliability as a software infrastructure problem. Complementary work
formalizes trust-scored geolocation as a persistent entity model for
Earth observation systems~\cite{padilla2026trust}, operationalizes
event-driven trust validation for rural autonomous
routing~\cite{padilla2026trustgeo}, deploys the architecture as an
IoT platform for smart city applications~\cite{padilla2026iot}, and
examines the governance conditions under which community-grounded place
representations support formal civic address
assignment~\cite{padilla2026reliable}. These works are cited for
architectural context; the round-trip offset oracle is independent and
can be adopted without reliance on any of them. Together, these works
address the full stack: detecting silent failures at the SE layer (this
paper), modeling longitudinal reliability, deploying trust-aware
infrastructure, and governing the addressing gap that makes the
detection problem necessary in the first place. The unifying goal
across these works is Reliable Places: a geolocation knowledge layer
in which operationally validated locations accumulate trust through
use, so that systems serving the same households, roads, or structures
do not have to rediscover their reliability from zero.

\section*{Data Availability Statement}
\label{sec:das}

The round-trip offset oracle pipeline---including the five-category
classification schema, HERE \resulttype correction logic, offset
computation, and the Algorithm~\ref{alg:rto} implementation---is
released as a self-contained package with a permanent DOI:
\url{https://doi.org/10.6084/m9.figshare.31868773}. The package
includes a synthetic dataset of 500 GPS coordinates in CSV format
(fields: \texttt{id}, \texttt{geolat}, \texttt{geolong},
\texttt{area\_type}) designed to exercise the oracle pipeline
end-to-end, enabling researchers to verify its behavior without access
to the deployment dataset. The oracle can be applied directly to any
GPS coordinate corpus without modification.

The deployment dataset is not released. It derives from a house-to-house
food assistance program serving approximately 35,000 households in a
single municipality during a public health emergency. Even with
coordinate-level anonymization, the combination of scale, geographic
specificity, and program context creates re-identification risks the
author is not able to fully mitigate---particularly for households in
sparse rural zones, where standard randomization radii may not be
sufficient to prevent parcel-level inference.

Readers can verify the oracle methodology and pipeline implementation
from the released code. The quantitative results in
Tables~\ref{tab:resolution}--\ref{tab:roundtrip} depend on the
deployment dataset and cannot be independently recomputed without it;
the ground truth validation in Section~\ref{sec:setup} documents the
methodology used to confirm those results against Puerto Rico's public
CRIM property registry.

\begin{acks}
This work was supported in part by the U.S.\ National Science Foundation
(NSF) under SBIR Phase I Grant No.\ 1913502. This material is also
based upon work supported by the National Science Foundation CISE
Graduate Fellowships under Grant No.\ 2313998. Any opinions, findings,
and conclusions or recommendations expressed in this material are those
of the author(s) and do not necessarily reflect the views of the
National Science Foundation.

English is not the author's first language. Claude (Anthropic) was
utilized to assist in drafting and revising text, formatting tables,
and developing the oracle pipeline implementation described in
Section~\ref{sec:das} in this work. All AI-assisted content was
reviewed, validated, and approved by the author, who takes full
responsibility for the accuracy and integrity of the work.
\end{acks}

\balance
\bibliographystyle{ACM-Reference-Format}
\bibliography{wyc_references}

\end{document}